# Hard X-ray polarimetry with Caliste, a high performance CdTe based imaging spectrometer

S. Antier[1] - P. Ferrando[1] - O. Limousin[1] - E. Caroli[2] - R. M. Curado da Silva[3] - C. Blondel[1] - R. Chipaux[5] - V. Honkimaki[6] - B. Horeau[1] - P. Laurent[1] - J.M. Maia[4] - A. Meuris[1] - S. Del Sordo[7] - J.B. Stephen[2].

**Abstract** Since the initial exploration of the X- and soft γ-ray sky in the 60's, high-energy celestial sources have been mainly characterized through imaging, spectroscopy and timing analysis. Despite tremendous progress in the field, the radiation mechanisms at work in sources such as neutrons stars, black holes, and Active Galactic Nuclei are still unclear. The polarization state of the radiation is an observational parameter which brings key additional information about the physical processes in these high energy sources, allowing the discrimination between competing models which may otherwise all be consistent with other types of measurement. This is why most of the projects for the next generation of space missions covering the few tens of keV to the MeV region require a polarization measurement capability. A key element enabling this capability, in this energy range, is a detector system allowing the identification and characterization of Compton interactions as they are the main process at play. The compact hard X-ray imaging spectrometer module, developed in CEA with the generic name of "Caliste" module, is such a detector.

In this paper, we present experimental results for two types of Caliste-256 modules, one based on a CdTe crystal, the other one on a CdZnTe crystal, which have been exposed to linearly polarized beams at the European Synchrotron Radiation Facility (ESRF). These results, obtained at 200 and 300 keV, demonstrate the capability of these modules to detect Compton events and to give an accurate determination of the polarization parameters (polarization angle and fraction) of the incoming beam. For example, applying an optimized selection to our data set, equivalent to select 90 degrees Compton scattered interactions in the detector plane, we find a modulation factor Q of $0.78 \pm 0.06$ in the 200 – 300 keV range. The polarization angle and fraction are derived with accuracies of approximately 1° and 5 % respectively for both CdZnTe and CdTe crystals. The modulation factor remains larger than 0.4 when essentially no selection is made at all on the data.

We also present in this paper a simple analytical model of the interactions for the detector geometry. We show that the experimental data compare very well with the simulation, and that simple geometrical effects explain some of the observed deviations between the data and the simulation.

All of these results, both experimental and from simulations, prove that the Caliste-256 modules have performances allowing them to be excellent candidates as detectors with polarimetric capabilities, in particular for future space missions.

**Keywords** Scattering Compton · Polarimetry · Schottky CdTe · CZT · Pixel detectors · Spectroscopy · Hard X-ray Astrophysics.

[1] S. Antier , C. Blondel , P. Ferrando, B. Horeau, P. Laurent, O. Limousin and A. Meuris
CEA Saclay, DSM/Irfu/Service d'Astrophysique
F-91191 Gif-sur-Yvette Cedex, France
e-mail: sarah.antier@cea.fr

[2] E. Caroli and J.B. Stephen
INAF/IASF-Bologna
Via Gobetti 101
I-40129 Bologna, Italy

[3] R. M. Curado da Silva
LIP-Coimbra, Departamento de Física, Universidade de Coimbra
PT-3004-516 Coimbra, Portugal

[4] J.M. Maia
Physics Department, University of Beira-Interior
6201-001 Covilhã, Portugal and LIP-Coimbra, Physics Department
University of Coimbra 3004-516 Coimbra, Portugal

[5] R. Chipaux
CEA Saclay, DSM/Irfu/Service d'Électronique, des Détecteurs et d'Informatique
F-91191 Gif-sur-Yvette Cedex, France

[6] V. Honkimaki
ESRF
6 Rue J. Horowitz, BP 220
F-38043 Grenoble Cedex 9, France

[7] S. del Sordo
INAF/IASF-Palermo
Via Ugo La Malfa 153
I-90146 Palermo, Italy

# 1 Introduction

Astrophysical studies of the γ- and X-ray universe are currently mainly based on imaging, spectral and timing analysis of high energy sources. Even though progress has been remarkable in these domains, thanks in particular to large dedicated space telescopes (such as Chandra, XMM, RXTE, INTEGRAL, SWIFT and Fermi); there still remain unsolved questions concerning high energy sources' physical processes. Source radiation emission polarization state (i.e. angle and degree of linear polarization) provide additional key information which can be used to address these questions and to distinguish between different models which are otherwise consistent when considering other source characteristics. In particular, polarimetric measurements provide information on the emission mechanisms at play, the emission region geometry, and the structure of the magnetic field in a wide variety of hard X- and γ-ray sources such as pulsars, solar flares, active galactic nuclei, galactic black holes or γ-ray bursts (Lei et al. 1997, Weisskopf et al. 2009).

Several theoretical models for these types of high energy sources predict the emission of fluxes with different levels of linear polarization, ranging from a few percent for AGN to several tens of percent for pulsars (Harding 2010, Matt 2010, Krawczynski et al. 2011). The complexity of making polarization measurements, coupled with the limited number of photons in the hard X- and γ-ray range have been the limiting factors that have kept this field practically unexplored in space astrophysics. The only missions with dedicated polarimeters ever flown are OSO-8 in the soft X-rays domain, with a successful measurement of the Crab nebula polarization (Weisskopf et al. 1976, 1978), and IKAROS, the small solar power sail demonstrator launched in 2010, in the gamma ray domain which carries the Gamma-Ray Burst Polarimeter (Yonetoku et al. 2011). Finally, in the gamma-ray domain, complex data analysis of the imaging spectrometers SPI and IBIS onboard INTEGRAL, have allowed to obtain pioneering on the Crab pulsar and Cygnus X-1 (Dean et al. 2008, Forot et al. 2008, Laurent et al. 2011, Chauvin et al. 2013), as well as on different GRBs (Coburn and Boggs 2003, Rutledge and Fox 2004, Kalemci et al. 2007, Götz et al. 2009, 2014) The interest in this field has generated a large number of proposals, advanced design projects, or even built but never flown instruments dedicated to polarimetry, and using non focusing optics. The will not be reviewed here as this is beyond the scope of this paper, but the interested reader is referred to Sofitta et al. 2013.

Still in the context of the instrumental work presented here, another path is now also being followed based on the progress made in focusing optics in the hard X- and soft γ-rays domain. This allows designing missions with an increase of several orders of magnitude in sensitivity and angular resolution with respect to non-focusing telescopes. The NuSTAR mission (Harrison, 2010) is the first flying mission of this type, and other missions with enhanced performances or different energy domains have been proposed e.g COSPIX (Ferrando et al. 2010), PheniX (Roques et al. 2012), GRI (Knödlseder et al. 2009), and DUAL (von Ballmoos et al. 2012). In these focusing telescopes, the focal plane has a small size, and the detectors need to have a very good spatial resolution to ensure that the imaging performance satisfies the requirements on angular resolution, and a very good spectral resolution to deal with requirements rising from astrophysical questions for which a clear identification of lines is required. Devices fulfilling these requirements are imaging spectrometers based on segmented and pixelated detectors. Provided that proper coincidence logic is used to read out the pixelated sensor, then the sensitive volume can also be used as a scattering polarimeter, taking advantage of the Compton interaction properties. This detector configuration allows optimizing the trade-off between the scientific return and the instrument design complexity thanks to the development of a single focal plane array playing all the roles. This is with the main prospect of such a telescope configuration for a space mission that we are presenting the results on the Caliste-256 modules in this paper.

In the hard X-ray / soft gamma-ray domain, the use of high Z material is mandatory, and in order to optimize a focal plane instrument for the different missions envisioned, CdZnTe polarimeters prototypes have already been tested in a series of experiments (Curado da Silva et al. 2004, 2008, 2012), which allowed the evaluation of their polarimetric performance and the study of possible sources of errors and factors limiting the achievable performance.

As a further step for improving the performance of this type of scattering polarimeter, we have decided to study the use of the fine pitch (580 μm), high energy resolution (~1 keV FWHM at 60 keV) and good time resolution imaging spectrometer developed at CEA, the Caliste-256 module (Limousin et al. 2011). In this paper, we quantitatively investigate such prototype detector performance, both from measurements performed under a linearly polarized beam at the European Synchrotron Radiation Facility (ESRF) in Grenoble (France), and from simulations dedicated to this detector configuration.

This paper is organized as follows. We first recall the measurement principle and describe the experimental setup configuration at the ESRF. We then detail the analytical model which has been specifically developed in this context, and which is used to generate simulation data in the same detector configuration as for the ESRF experiment. The next section presents the spectral analysis of the experimental data, including corrections for charge splitting and escapes lines. This is then followed by the description of the polarization analysis procedure, common to both experimental and simulation data, and to the study of systematics effects. In the last section, before the conclusion, are presented all the experimental results obtained in this experiment, and its comparison with simulation data.

## 2  Scattering polarimetry principle

The possibility to measure polarization with spectro-imaging detectors such as that we are developing relies on the fundamental physics underlying the Compton scattering process for linearly polarized photons. This process presents a dependence on polarization direction given by the Klein-Nishina (1929) differential cross-section for Compton scattering per unit of elementary solid angle $d\Omega$:

$$\frac{d\sigma}{d\Omega} = \frac{r_0^2}{2} \left(\frac{E'}{E}\right)^2 \left[\frac{E'}{E} + \frac{E}{E'} - 2\sin^2\theta \cos^2\varphi\right]$$

(Eq. 1a),

where $r_0$ is the classical electron radius and $E$, $E'$ are the energies of the incoming and scattered photons respectively. The scattered photon is deviated from its original direction by $\theta$.

$E$, $E'$ and $\theta$ are related by the following equation:

$$E' = \frac{E}{1 + \frac{E}{m_e c^2}(1 - \cos\theta)}$$

(Eq. 1b),

with $m_e c^2$ being the electron rest energy.

The azimuthal deviation angle $\varphi$ corresponds to the angle formed by the scattering plane (defined by the initial direction and the scattered direction, see Figure 1a) and the incoming photon polarization plane (defined by the photon direction and its polarization vector, i.e. the photon electric field component).
Considering linearly polarized photons, Eq. (1a) provides the azimuthal ($\varphi$) dependency for the Compton scattered photons. The Compton scattering of polarized photons generates non-uniformity in its azimuthal angular distribution (Figure 1b). From the equation, fixing all parameters except the azimuthal angle $\varphi$, the probability of interaction reaches its minimum and maximum for orthogonal directions, $\varphi = 0°$ and $\varphi = 90°$ respectively.

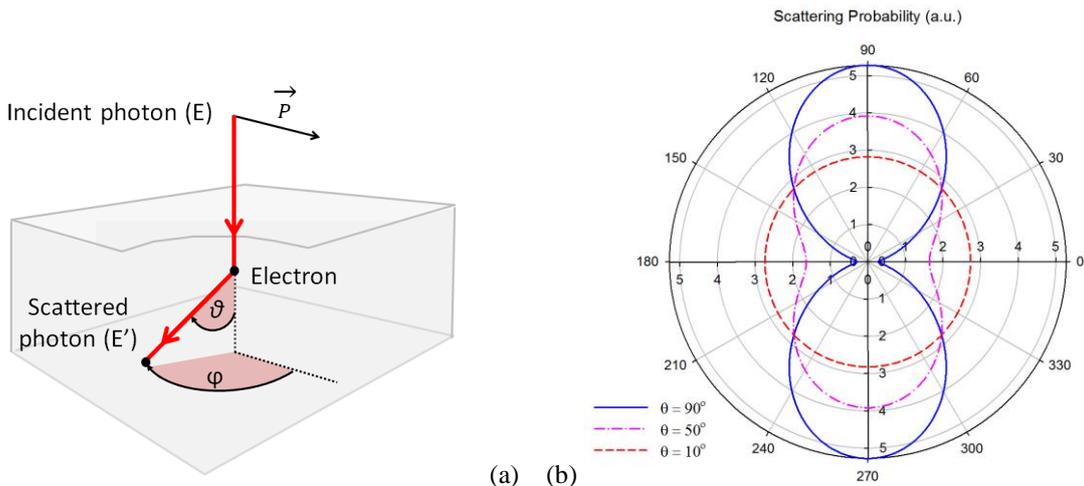

Figure 1. (a) Scheme of Compton scattering of a polarized photon; (b) Azimuthal angle ($\varphi$) probability distribution for a given Compton scattering angle ($\theta$) of linearly polarized photons at $E = 200$ keV. The direction of the polarization is parallel to the horizontal axis of the polar plot.

The polar plot in Fig. 1b shows the probability distribution against the azimuthal angle $\varphi$ for linearly polarized photons at 200 keV, after Compton scatterings at $\theta$ angles of 10, 50 and 90°. It is clearly apparent that the asymmetry of this distribution increases with the scattering angle. Note that the asymmetry is almost invisible at $\theta = 10°$, while for $\theta = \theta_M = 90°$, the probability of a photon to scatter perpendicularly to the polarization plane is roughly 5 times larger than that to scatter along the polarization plane, with the probability distribution looking like a bowtie (cf. Figure 1 and Figure 5).

The measurement of polarization consists in measuring, and characterizing the shape of this distribution. To do this, one has to identify both the location of the Compton interaction (the center of the bowtie distribution), and the direction of the scattered photon. This latter information can be obtained if the scattered photon interacts and this interaction location is measured. In the range of energies we are testing, this second interaction is usually a photoelectric absorption. A polarimetric measurement thus necessitates the measurement of two simultaneous events, one at the location of the Compton interaction, with an energy deposit in the pixel $E_d$, and one at the location of the photoelectric absorption with an energy deposit $E_p = E'$, with $E_d + E' = E$.

In practice, with a single thin detector, pixelated, the best configuration is for a perpendicular illumination of the detector, in the center of one of its central pixels. Photons scattered around $\theta = 90°$, for which the polarization information is maximum, can then be stopped in the same detector, in one of the peripheral pixels, thanks to the large amount of matter provided in this configuration. By analyzing the spatial distribution of this second interaction position, for double events, which has the imprint of the bowtie shape, one can derive the degree of polarization and the polarization direction of the incident radiation.

At this point, it is possible to define the widely used figure of merit for a polarimeter: the polarimetric modulation factor $Q$. For a pixelated detector, $Q$ is written as:

$$Q = \frac{1}{P} \frac{Nmax - Nmin}{Nmax + Nmin} \qquad \text{(Eq 2.)},$$

where $N_{max}$ and $N_{min}$, are the maximum and minimum of the angular azimuthal distribution of the scattered photons defined over the detector plane and P is the degree of polarization. Because of the nature of the scattering process, $N_{max}$ and $N_{min}$ are counted along two orthogonal directions.

## 3 Experimental Setup

As explained in the previous section, the fraction and the angle of polarization of the incoming radiation are determined by measuring the azimuthal angular distribution of the scattered photons after the impinging photons Compton scatter into a position sensitive spectrometer detector: double-hit positions, arrival time and energy deposits are measured simultaneously and independently. The polarization of the source is derived from the data after a sufficient accumulation of events.

In May 2011, with the aim of evaluating the performance of CZT/CdTe based room temperature solid state pixel spectrometers as hard X ray scattering polarimeters, our collaboration set up an experiment at European Synchrotron Radiation Facility (ESRF Grenoble, France) using the ID15A high energy beam line: POLCaliste, i.e POLarisation with Caliste-256.

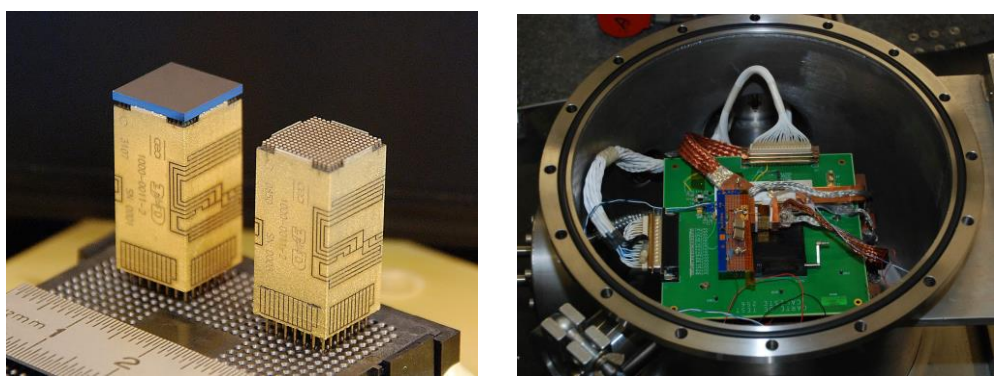

Figure 2. (a) The 1 mm thick Schottky CdTe Caliste-256 module with 580 μm pixel pitch; (b) The Caliste module mounted on its electrical and mechanical interface inside the vacuum container (T ≈ -10°C).

### 3.1 The Caliste-256 detection module

In order to measure the two interactions, one corresponding to a Compton scatter, the other to the absorption of the scattered photon, a fine pixel and high spectral resolution detector, named Caliste-256 was employed (Limousin et al 2011). Caliste-256 is a hybrid camera that integrates a 1 mm thick CdTe Schottky (or a 2 mm thick CZT) monolithic crystal pixelated detector mounted on top of a 3D module supporting the front-end electronics based on 8 full custom ASICs IDeF-X (Gevin et al. 2006) developed at CEA (Figure 2). This detector module is able to detect photons between 2 keV and 280 keV. With a 580 μm pitch, arranged in a 16 × 16 pixels array, and 1 keV (FWHM) energy resolution at 60 keV, the Caliste-256 version has been specially developed for space missions taking into account environmental space constraints (Ferrando et al. 2005). Its characteristics enable a straightforward selection of Compton events, and a good determination of the geometry of the photon interactions.

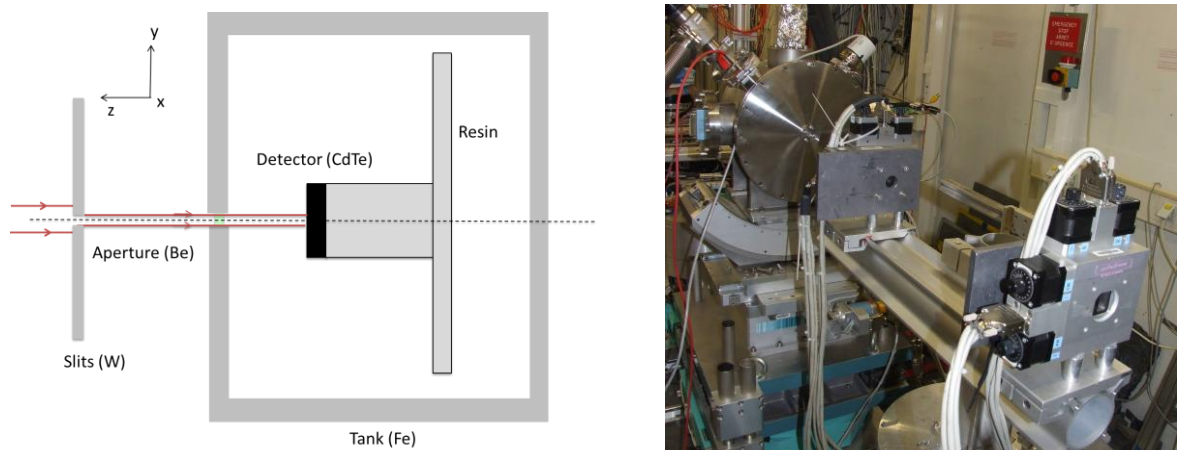

Figure 3. Schematics (a) and picture (b) of the PolCaliste experiment set-up at experimental hutch of the ESRF ID 15A beam line in Grenoble (France).

3.2    The ESRF experiment setup

The ID15 beam line at ESRF (Grenoble, France) provides high-energy and mono-energetic X-ray photon fluxes with a user selectable energy between 30 – 750 keV and a linearly polarized fraction up to about 100%. As can be seen in Figure 3a, the mono-energetic X-ray beam was directed onto the top surface of the Caliste-256 module, confined in a vacuum chamber, through a collimation system with adjustable tungsten slits. This collimator allowed the beam size to be set to 100 μm × 100 μm, significantly smaller than the 500 μm pixel size. We scanned the detector surface with beams of different energies in the dynamic range of Caliste-256 in order to perform a fine energy calibration of each individual pixel. We checked the detector response uniformity and evaluated the overall system performance prior to performing dedicated Compton runs for which we centered the beam spot onto one of the four central pixels.

Figure 3b shows, in the foreground, the X-ray beam collimation system, and in the background the flange of the vacuum vessel which has at its center a beryllium entrance window placed in front of the cathode side of the Caliste-256 module. The vacuum vessel hosts the Caliste-256 mounted on its test board, used to supply the voltages and data link and is cooled down to -10°C. The whole setup is installed on a movable system. The micrometric precision translation and rotation stages allow the accurate positioning of the detector in the plane (X, Y), perpendicular to the beam axis Z, and its rotation around the direction of the beam itself.

| Type of detector | CdTe | CZT |
|---|---|---|
| Thickness (mm) | 1 | 2 |
| Incident beam Energy (keV) | 200-300 ||
| Degree of polarisation (%) | 80-85-88-89-90-98 ||
| Angle of polarisation (degrees) | 0-5-10-20-30 ||
| Model analysis | Yes ||

Table 1 – Parameters tested with Caliste-256

Table 1 summarizes the set of measurements, which have been performed for both Caliste-256 samples (1 mm CdTe and 2 mm CZT) under different conditions: beam energies, degree and angle of polarization. Approximately 1 million events were recorded for each run defined by a unique experimental configuration. The data are time tagged photon-photon lists of events, which were then analyzed off-line: energy calibration, double event selection, spectra build up.

3.3    The Caliste-256 triggering logic for the ESRF experiment

Caliste-256 is an imaging spectrometer with fine timing resolution. Each of the 256 pixels is independent from the others. Each channel has its own preamplifier, pulse shaper, adjustable discriminator, and its own peak detector system. Note that Caliste-256 has no pile-up rejection system. Caliste-256 has no sample and hold circuitry but a peak detector (the peak detector records the maximum of the amplitude at each shaper output). The detailed architecture is described in Limousin et al 2011. Each of the eight IDeF-X ASICs inside Caliste-256 shares for all its 32 channels a common hit pixel register, an analogue memory, a multiplexing analogue output system, a slow control digital link and a triggering system. All discriminator outputs are routed to an OR logic system to deliver a unique trigger signal for one ASIC. Inside Caliste-256, the eight ASIC trigger outputs are routed again to a cabled OR logic. Thus Caliste-256 has a unique global trigger for the whole camera.

When a photon hits the CZT sensor and deposits an energy above the discrimination threshold, a trigger signal is generated and the address of the hit pixel is stored into the hit pixel register of the associated ASIC. The Caliste-256 global trigger signal rises when at least one channel of the camera has been hit. The Caliste-256 global trigger signal is sent to the FPGA controller to enable the readout sequence. The trigger is time tagged with a resolution of 20 ns. Note that the trigger resolution is not the sensor resolution, which is in the range of one microsecond. After a 10 µs latency time, the delay required to develop the pulse shape up to its maximum, Caliste-256 is locked. The latency time is set to 10 µs, which corresponds to the shaper peaking time value, enabling accurate pulse height measurement, free of ballistic deficit. During the latency time, pile-up may occur when the count rate is high. All peak detector outputs are held and the corresponding amplitudes are stored into each channel analog memory. Any new upcoming photon hit will be ignored. Once locked, Caliste-256 is ready to communicate with its FPGA controller to release the memorized data. At first, the FPGA reads the hit register of each ASIC to determine the addresses of the hit channels. It computes and starts the multiplexing sequence to present the analogue data of the corresponding channels to the system ADCs. The data packet is created as a list of events. The latter contains the global trigger date, hit pixel addresses and amplitude binary codes. The data packet is transmitted to the Data Acquisition System via a Space-Wire link. Because the channels are read in serial mode, the readout dead time is dependent on the number of hit channels. The typical value is 17 µs plus 1 extra µs per channel to be read. When the readout sequence is completed, Caliste-256 is reset and the acquisition resumes.

For a Compton event, two hits arise into two distinct channels. In this case, one or two ASICs trigger exactly at the same moment but a single global trigger is set up, as depicted above. This event generates, for a unique time tag corresponding to the global trigger, a set of two addresses and two corresponding energies. The data packet has a list of two events having precisely the same date. Similarly, a photon hit accompanied with a photon escape in a neighboring pixel generates a similar kind of event in the list. Conversely, when two independent events hit the sensor in the duration of the latency time, the first of the two raises the trigger. Consequently, there is a chance that the second event energy is not recorded properly as its pulse height might not reach its maximum before being read out. The probability of such an event is related to the detector count rate and the 10 µs duration of the latency time, which can be seen as a coincidence window. At ESRF, where the beam was targeted onto a single pixel at a time, this probability was very low apart in the targeted pixel itself. Compton scattered energy in surrounding pixels is very likely to be measured accurately as shown in further.

Thanks to the architecture of the Caliste-256 electronics, it is possible to measure simultaneously the energy deposits in different pixels from the same impinging photon after it has Compton scattered. Consequently, the selection of Compton events in the photon list is straightforward, which facilitates an accurate construction of the bowtie figure.

The Caliste-256 trigger is very flexible. Due to the low probability of recording a Compton double event in the detector volume (~5% of all events at 200 keV beam energy) we chose to set the trigger criterion so as to record only double and more hits in order not to saturate the telemetry with simply useless data. On the contrary, the rejection of the single events prevents us from measuring the Compton to photoelectric probability. The way to adjust the trigger system is as follows: The Caliste-256 global trigger works in current mode. In other words, the global trigger signal is the sum of the ASIC trigger signal currents. For instance, the trigger signal amplitude out of Caliste-256 when two chips are hit simultaneously, in at least one pixel each, is exactly twice the signal amplitude obtained with one single chip trigger. Discriminating on the trigger height forces the FPGA controller reaction only when at least two ASICs are touched. The drawback of this configuration is that there is no trigger generated for single interactions, in particular in the beam pixel and for Compton double hit when the two corresponding pixels are read by the same ASIC (two rows). The readout happens only when a second chip is hit. However, apart the central pixel information, no information is lost because the signal amplitude is kept in the chip memory until the next reset: all scattered photons interacting into the detector are accurately transferred (energy and position of interaction) to the acquisition system but the corresponding primary energy deposit into the central pixel is lost as it is almost immediately followed by a photon at the beam energy: due to the peak detector behavior which records the maximum signal in the latency time, the central pixel energy is almost always the beam energy. We checked that the bowtie figure is not

## 4 Numerical simulation of the polarimetric performance

In order to support the interpretation of the experimental results, and to assess the reliability of the Caliste-256 module polarimetric performance, we have developed a 3D numerical tool to calculate the expected polarization performance in the energy range of interest here, from 50 to 300 keV. Based on tabulated photon-matter interaction probabilities, the model allows the theoretical azimuthal distribution of Compton scattered photons to be constructed. The input parameters of this model are, on the detector side, the crystal composition (CdTe and CZT) and its geometry (number of pixels, pixel pitch, inter-pixel gap and detector thickness), and, on the incident photon beam side, its energy, its position and angle of incidence with respect to the surface of the sensor, and its degree and angle of polarization.

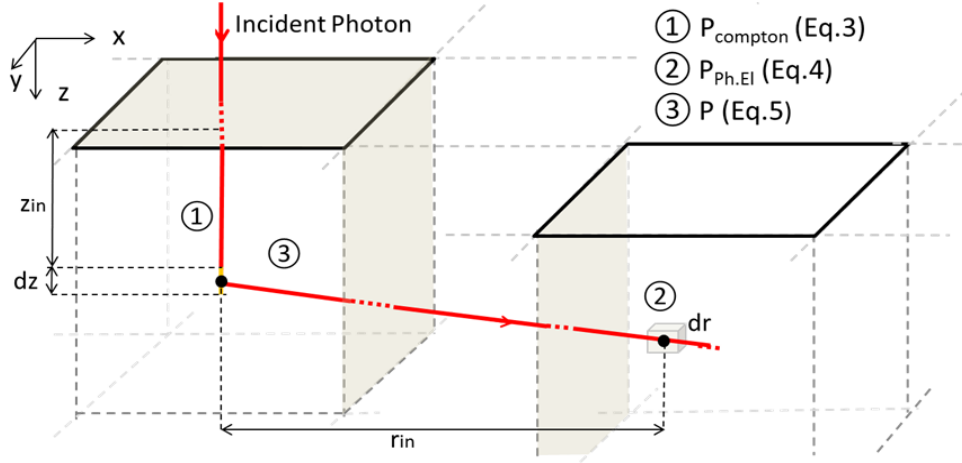

Figure 4. The elemental process implemented in the simulation tool, with proper integration on the detector thickness and pixel location and volume.

The model, implemented in IDL (Interactive Data Language), is based on the analytical expressions describing the following elemental process depicted in Figure 4: a Compton interaction at depth $z_{in}$ in the central pixel, with the scattered photon emitted with angles $\theta$ and $\varphi$, followed by a photoelectric absorption of this scattered photon after travelling a distance $r_{in}$ without any interaction. These are given by:

$$dP_{Compton}(z_{in}, E) = e^{-\mu_T(E) \cdot \rho \cdot z_{in}} \cdot \mu_C(E) \cdot \rho \cdot dz \qquad \text{(Eq. 3)},$$

$$dP_{Ph.El.}(r_{in}, E') = e^{-\mu_T(E') \cdot \rho \cdot r_{in}} \cdot \mu_{Ph.El.}(E') \cdot \rho \cdot dr \qquad \text{(Eq. 4)},$$

$$P(\theta, \varphi) = \frac{1}{\sigma_{tot}} \frac{d\sigma}{d\Omega} \qquad \text{(Eq. 5)}.$$

where $\mu_C$, $\mu_{Ph.El.}$, and $\mu_T$ are respectively the Compton, photoelectric and total mass attenuation coefficients, $\rho$ is the density of the detector material, $E$ is the incident photon energy and $E'$ that of the scattered photon, and $d\sigma/d\Omega$ is the Klein Nishina differential cross section given in Eq. (1) with $\sigma_{tot}$ its integral over the entire solid angle.

These equations are integrated over the relevant parameters for detector thickness and pixel location and volume. For a given Compton interaction position ($z_{in}$) and scattering direction ($\theta$ and $\varphi$) the integration over $r_{in}$ is performed analytically for each pixel. The integration over $z_{in}$, $\theta$ and $\varphi$ can be performed only numerically, which is performed with sufficiently small steps with respect to the interaction length ($z_{in}$) or the characteristic variation scales ($\theta$ and $\varphi$).

This model does not take into account more complex phenomena, as e.g. the Doppler broadening of the Compton interaction, or the energy resolution of the detector, or processes which could give rise to higher multiplicity events (such as a second Compton interaction or an escape line). This would necessitate a full Monte-Carlo simulation. However, as will be seen in the following sections, this analytical model is more than accurate for our purposes. It describes our observations very well, and allows simple and fast investigations regarding the different beam and detector parameters.

Figure 5 shows the theoretical distribution of double events generated by this model, a Compton scattering in the central pixel followed by a photoelectric absorption of the scattered photon in a peripheral pixel. In Figure 5a, the distribution showing a bowtie shape is obtained by simulating a 100% linearly polarized beam at 200 keV, normal to the detector surface with an angle of polarization of 30°. In this case, the main scattering axis of distribution is perpendicular to the polarization vector. In Figure 5b, the same simulation is performed but the impinging flux polarization fraction is set to 0. In this case, no preferred direction is visible on the map.

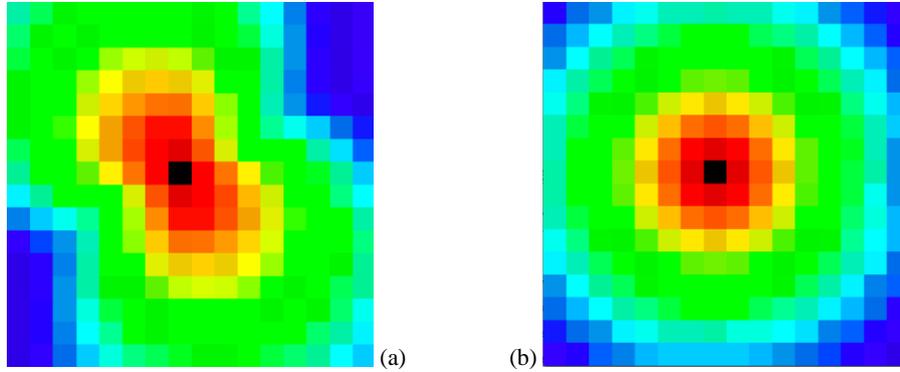

(a)                    (b)

Figure 5. Compton scattered photons distribution for a 200 keV incident energy beam predicted by the numerical model. The color code corresponds to the probability: Red is high while blue is faint. (a) 100% linearly polarized beam with angle of polarization = 30°; (b) un-polarized photon beam.

## 5  Experimental data spectroscopic analysis

As the Compton scattered photon energy is related to its angle of deviation, a careful energy calibration and a detailed analysis of the observed spectral lines is mandatory before attempting to derive the Compton polarimeter performances. The typical spectra in the central pixel and the surroundings pixels are displayed in Figure 6. They show complex features, and in particular include two phenomena that can be corrected for: fluorescence escape lines and charge sharing between adjacent pixels. These spectral analysis and corrections are described below.

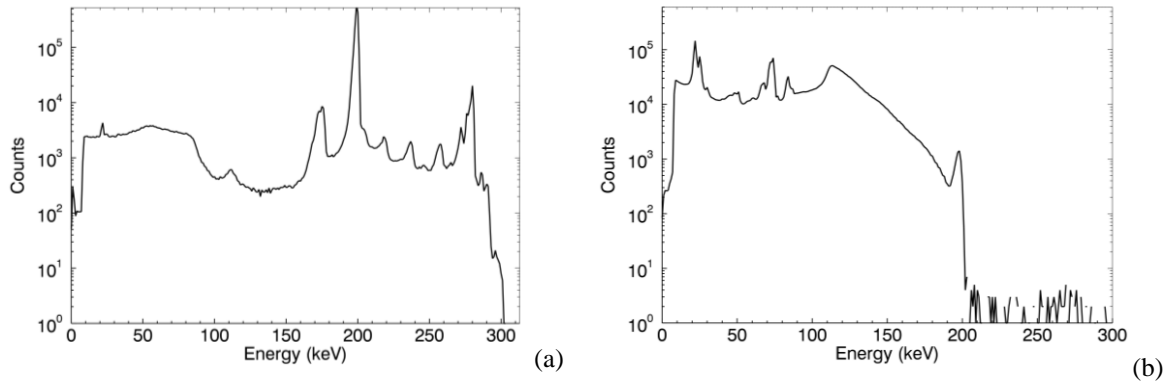

(a)                    (b)

Figure 6. Energy spectra of CdTe Caliste-256 detector, irradiated by 200 keV photons, (30° angle of polarization and 98% degree of polarization) (a) spectrum of the central pixel (b) spectrum of the all pixels except for the 9 at the center.

We focus here on the analysis of spectra obtained at 200 keV with a 1 mm thick CdTe sample but an identical analysis has been performed at all beam energies and for both CdTe and CZT detectors. In order to achieve the full detector energy calibration, we scanned the surface with the beam, pixel by pixel. Figure 6a shows a spectrum from the target (central) pixel, after pulse-height calibration, while Figure 6b is the spectrum from the sum of all the pixels of the detector, except for the central pixel and the 8 adjacent ones (the "9 central pixels").

In the central pixel (Fig. 6a), where the beam is aimed, apart from the photoelectric peak at precisely 200 keV we can identify the following expected structures:

- Fluorescence lines from matter surrounding the detector, mainly W ($K_\alpha \approx 59$ keV and $K_\beta \approx 67$ keV) and Pb ($K_\alpha \approx 75$ keV and $K_\beta \approx 85$ keV). W is present in the slits located at the front of the detector, which are used to fine shape the beam. Lead is massively present on the ID15A hutch walls, facing the detector and used as shielding. Due to their position with respect to Caliste-256, the W and Pb fluorescence photons are rather uniformly distributed over the CdTe surface.
- The Compton edge at 87.8 keV.
- The backscatter peak at 112 keV, which is due to the interaction of photons with the electronic material behind the sensor.
- The Cd ($K_\alpha \approx 23$ keV) and Te ($K_\alpha \approx 27$ keV) fluorescence escape lines, respectively at 177 and 173 keV.
- Some combinations of piled-up events above 200 keV. Note that Caliste-256 has no pile-up rejection system and has no sample-and-hold circuitry but a peak detector. Thus, Caliste-256 is sensitive to pile-up when several photons interact into the same pixel during the latency time (time elapsed between the trigger and the internal hold signal commended 10 µs later by the FPGA sequencer).
- The saturation of the analogue front-end electronics for pulse heights approximately corresponds to 270 keV. The saturation is not sharp due to the readout chip design. The shape of the saturation is complex because the dynamic range of our readout electronics depends on the count rate. Usually, saturation appears as a peak ending the dynamic but it looks different in Caliste-256. One should consider that Caliste-256 response is no longer valid above 270 keV and events above must be ignored. However, such events are not discarded in our data set as they contribute to the dead time.

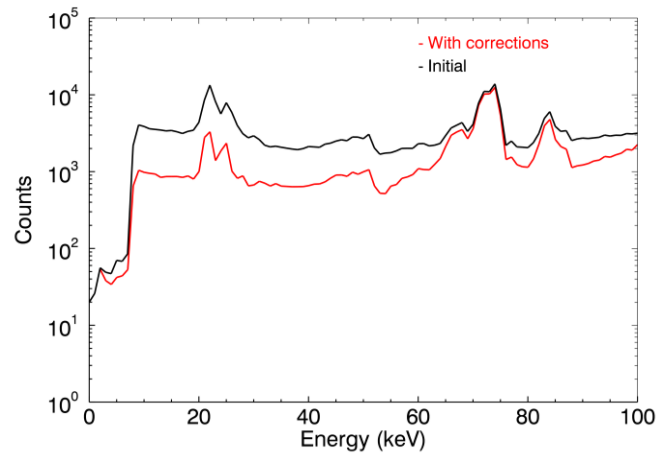

Figure 7. Initial and corrected energy spectrum of the pixels around the 9 central pixels of CdTe Caliste-256 detector, irradiated by 200 keV photons, (30° angle of polarization and 98% degree of polarization)

Regarding now the pixels around the central one (Fig. 6b), we identify the following expected structures:
- The Cd ($K_\alpha \approx 23$ keV) and Te ($K_\alpha \approx 27$ keV) fluorescence lines.
- The surrounding material fluorescence lines, mainly W ($K_\alpha \approx 59$ keV and $K_\beta \approx 67$ keV) and Pb ($K_\alpha \approx 75$ keV and $K_\beta \approx 85$ keV) as described previously.
- The Compton scattered photons from 112.2 keV up to the photo peak energy with a particular population of interest around 143.8 keV corresponding to the Compton scattered photons at exactly 90° with respect to the beam direction.
- The bump at 195 keV up to the photo peak energy of Compton photons scattering in front of the detector (probably with the beryllium window), corresponding a scattered angle between 0 and 20°.
- A continuous spectrum due to charge sharing between neighboring pixels. This type of event occurs when a photon hits the detector in between two pixels or when the charge spreads out a pixel and induces a fraction of the signal on a neighbor pixel.

Prior to build the modulation curve and to analyze the polarimetric performances of Caliste-256, we apply spectral corrections for split events (charge sharing) and CdTe X-ray fluorescence escapes. Such events are responsible for incorrect energy measurement in a pixel where a Compton scattered photon has interacted.
In case of Cd or Te X-ray fluorescence photon escape in neighbor pixels, the sum of the energies is reassigned to the primary pixel position.
Besides the fluorescence, 21% of the Compton scattered photons have theoretically a chance to interact in between two pixels generating a charge sharing (Iniewski 2007). In our case (CdTe, 1mm, 300V -10°C, 2 keV threshold) we measured a probability of 19.8% to record a split event. This effect is corrected as well, by summing neighboring double events. The computing process is very similar to the X-ray fluorescence apart the energy distribution is continuous. We chose to reassign the location of the event to the pixel position where the highest charge has been detected. Finer corrections can be engineered. In the energy band from 19 to 31 keV, combining fluorescence and charge sharing, up 76% of the photons are reassigned to a neighbor pixel.
This correction increases the Compton hump in the range from 112 up to 195 keV by more than 4%. The remaining ~24% uncorrected fluorescence and charge sharing events are essentially due to isolated events located at the vicinity of the guard ring. The 200 μm guard ring that surrounds the pixel array is not connected to a spectroscopic channel. Thus, a photon escape or split event with the guard ring cannot be reassigned to any pixel. The effect of the correction is illustrated on figure 7.

## 6 Scattering maps, modulation curve construction and polarization parameters determination

We build the modulation curve searching for the azimuthal distribution of Compton scattered events in the Caliste-256 sensor plane. A crude analysis, taking into account all the photons detected in the pixels surrounding the central pixel (targeted by the X-ray beam), would result in a moderate modulation factor Q in the range of 0.4. Conversely, taking advantage of Caliste-256 fine resolution, spectral corrections as shown above and selection of events in narrow bands of energy are possible. We chose the energy bands to select Compton scattered photons strictly deviated by 90 degrees in the detector plane. This approach allows a maximization of the modulation factor Q, up to ~0,8, at the expense of a lower efficiency.

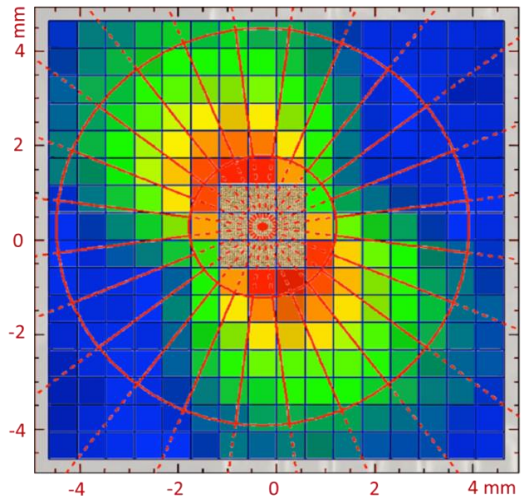

Figure 8. A false color map of the CdTe Caliste-256 counts distribution of events with scattered energies in the 139-148 keV band. These data are for a 200 keV photon beam, 98% linearly polarized and with 30° polarization angle. The 9 central pixels are off scale. On top of the data are shown the sectors, with their inner and outer radii, and their angular limits, which are used to count the events for building the azimuthal distribution curves.

In the following we chose to build the modulation curve by selecting the Compton scattered events corresponding to a scattering of the beam photons at the maximum modulation of around 90°. At 200 keV and 300 keV this corresponds to a scattered photon energy (which we will refer to as simply "energy" in the following) of 144 keV and 189 keV respectively. Accounting for Doppler broadening (Zoglauer and Kanbach 2003) and the detector energy resolution (1 keV FWHM), we decided to select events with energies in the 139 – 148 keV range for the 200 keV beam, and 181 – 197 keV range for the 300 keV beam. Note that in the case of an ideal detector and no Doppler broadening these energy ranges would theoretically correspond to a scattering at $90 \pm 6°$. This is the only selection, which is applied to the data. A map of the corresponding counts in these ranges reveals the expected bowtie shape (see Figure 8).

The final step of the data analysis is the extraction of the modulation curve. It consists of building the distribution of the Compton scattered photon count-rate with respect to the azimuthal angle. To do so, we superimpose a grid with 24 angular sectors, of 15° each, on the 2D Compton event distribution map as shown in Figure 8. The inner and outer radii of the grid are set to 1.5 and 4.2 mm respectively. The inner value is chosen to exclude the 9 central pixels, which give poor information on the angular direction of the scattered photons; the outer value is chosen so that the entire grid is included in the square shaped detector surface. The center of the grid coincides with the beam spot position on the central pixel.

The number of counts from a given pixel assigned to each sector is measured in the following way. If a pixel is fully included into the sector surface, the total number of counts of the pixel is added in the sector count. If only a part of the pixel is included in the sector, then the number of counts, fractional, assigned to that sector is given by the total number of counts of the pixel weighted by the fraction of the pixel surface, which intersects the sector. This first order correction of the pixel-size effect allows the construction of a smooth modulation curve as a function of angle, as shown on Figure 9.

The same method is applied to the data obtained from the simulation described in Section 4, which are organized as count maps, as for the experimental data. When compared to experimental data, the simulated data are normalized so that the total number of counts is the same for both sets. Figure 9 shows both experimental and simulated data for the case of a 300 keV, 98% polarized beam with a 0° polarization angle.

As expected, the first maximum is found at +90°. The modulation curve from the simulation is represented on Figure 9 in a sinusoidal full line. The model almost perfectly matches the data. The experimental Q factor is 0.73 ± 0.02.

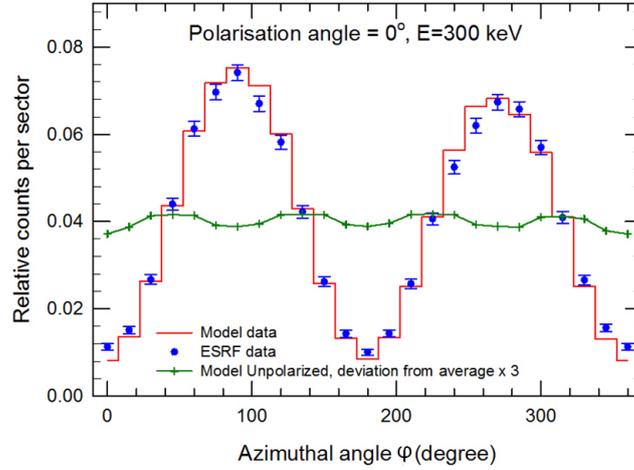

Figure 9. Modulation of scattered photons distribution for a 300 keV incident beam with a 98% linear polarization at an angle of 0°, for CdTe Caliste-256. A simulated modulation histogram is superimposed to the data taking into account the true incidence angle of the beam. The curve in green is the result of a simulation for a non-polarized irradiation: the fluctuations (with respect to average) are multiplied by 3.

Using the same simulation model, we have calculated the modulation curve for a non-polarized beam at the same energy, ie 300 keV. The corresponding modulation curve is pretty flat but not perfectly straight, as a perfect system would show. The response to a non-polarized beam is shown in Figure 9, intentionally magnified by a factor of 3 to help the reader to appreciate the shape of the parasitic modulation structure. This parasitic modulation is a systematic error due to the square shape of the pixels sampling the bowtie figure. The effect will be one of the limiting factors of our Minimum Detectable Polarization (MDP) (Weisskopf et al. 2009). Although the paper aims at studying of the polarimetric performances of Caliste-256 in laboratory conditions and the reader must be aware that others systematics errors may contribute to the loss of polarimetric sensitivity in a given telescope configuration. For example, the reconstruction of the modulation curve will be affected by the continuous spectrum of a gamma sources and the pretty uniform beam impinging over the detector surface. In this case, other strategies are employed to control the systematics such as discussed by Muleri and Campana (2012).

Finally, for fast fitting purposes of the experimental data, we have also used a simplified simulation of the modulation curve, in the 90° scattering conditions, based on the integration of Equation 1 over the azimuthal range corresponding to the sectors, with parameters being the polarization angle $\varphi_{obs}$ and the fraction of polarization $P_{obs}$. In this simplified model, the finite dimension of the pixels is not taken into account. Using the "curvefit" function of IDL, on can find the best parameters describing the data. In the case of Figure 9, they are $\varphi_{obs} = 0.09 \pm 0.29°$ and $P_{obs} = 90.4 \pm 0.7$. The results for all measurements are listed in the Table 2 of Section 7.

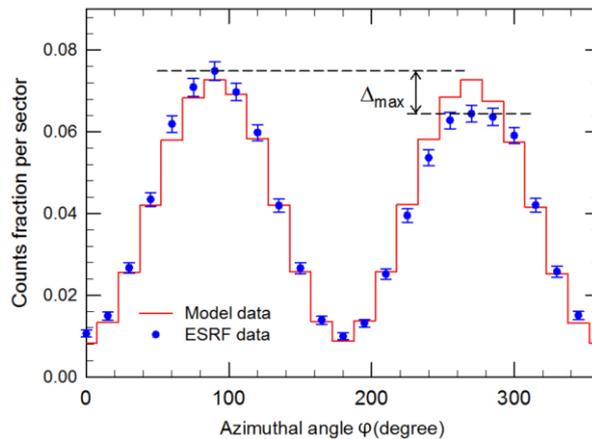

Figure 10. Modulation curve obtained from real data at 300 keV, 98% polarized beam, 0° polarization angle. Photons energy selection from 181 to 197 keV corresponds to 90 degree Compton scattered events only. A simulated modulation curve (histogram) is superimposed to the data and reveals an asymmetrical behavior when the beam is supposed to be strictly normal to the detector surface.

## 6.1 Effect due to the inclination of the detection plane

Looking at the modulation curve maxima on Figure 9, it is noticeable that the plot is asymmetric: the first peak is significantly higher than the second one. We call $\Delta_{max}$ the difference of the peak heights, as shown in Figure 10. The curve will only be fully symmetric, i.e. $\Delta_{max} = 0$, if the incident beam is perfectly perpendicular to the detector. Any tilt of the detector surface causes a non-uniformity in the Compton scattered photon range: selecting the energy of the Compton scattered photons corresponding to a 90° scatter changes the effective thickness of the detector in each azimuthal direction, because the scattered photon direction is no longer in the detector plane. Consequently, this impacts the modulation curve by shifting the relative heights of the maxima, i.e. $\Delta_{max} \neq 0$. The impact depends on the beam energy and the polarization angle. These results are consistent with the recent work published by Muleri et al. (2014), devoted to the analytical calculation of a Compton polarimeter response when incident photons are impinging on the detector plane off-axis.

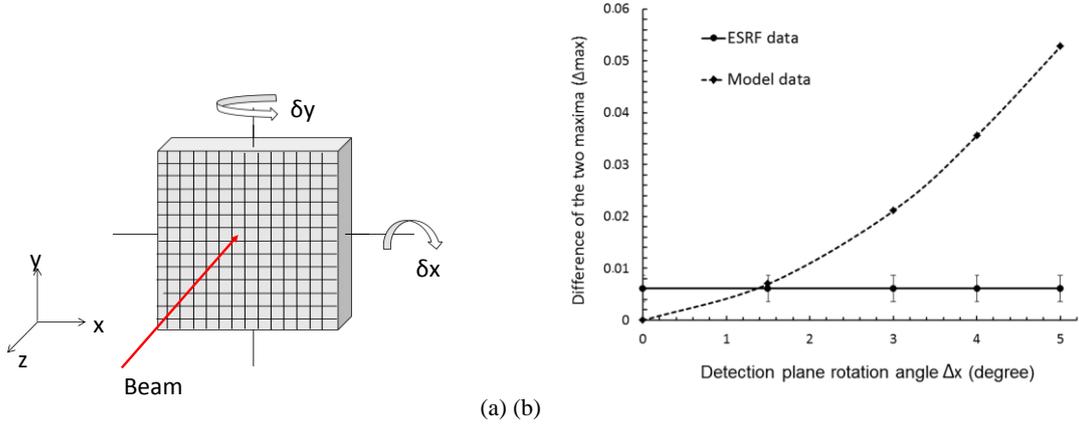

Figure 11. a) Representation of the two rotations angles $\delta x$ and $\delta y$; b) $\Delta_{max}$ against detector tilt along the x-axis. $\Delta_{max}$ is the absolute difference of maxima in the modulation curve, expressed in counts fraction per sector, for 24 sectors. A perfectly perpendicular incidence of the beam with respect to the detector surface gives $\Delta_{max} = 0$. The plot is calculated for 300 keV, 0° polarization angle. The model is compared to the data.

On Figure 10, the modulation curve model superimposed to the data assumes a perfectly normal incidence for the beam and reveals a deviation to the data at the maxima positions. $\Delta_{max}$ is about 10% of the modulation amplitude. Simulations have demonstrated that by tilting the detector surface with respect to the beam direction, one can account for this effect, as shown on Figure 9 for the same data set.

In our experimental setup conditions, it is extremely difficult to precisely control the detector orientation. The detector is inevitably tilted with respect to (zOy) and/or (zOx) planes by $\delta x$ and/or $\delta y$ rotation angles respectively, as defined on Figure 11a. The modulation curve symmetry is very sensitive both to the incidence angle and the polarization angle, as is shown in Figure 11b. Simulating a 0° polarization angle at 300 keV, we see that a rotation angle $\delta x$ along the x-axis in the range of ~1° will cause a maxima height separation by $\Delta_{max} \approx 0.006$ counts fraction/sector, i.e. ~ 10% of the modulation amplitude. The effect is non linear and increases with the tilt angle.

Fitting the angle $\delta x$ is a measurement of the detector surface position with respect to the beam direction. From Figure 11b, we find the detector surface tilt to be $\delta x = 1.5°$. In this case, the fitting of the angle $\delta y$ is not decisive because there are few counts recorded in the direction of polarization. The same analysis with 10/30°, 300 keV leads to a same fitting value of $\delta x = 1.5°$.

## 7 Full experimental results, and polarimetric response and sensitivity of Caliste-256

From the data treatment and analysis described above, we have characterized the response of the Caliste-256 as a polarization detector, depending on the beam energy, the degree and angle of polarization, and the type of crystal. For each run, after the spectral corrections described in Section 5, we have built the modulation curve using only an energy selection criterion, 139 – 148 keV at 200 keV and 181 – 197 keV at 300 keV. From this modulation curve, we have on the one hand measured the Q factor, which has been reached in this configuration, and on the other hand derived the best fit for the polarization parameters (angle, fraction), using the method explained in the previous section. Table 2 summarizes the results for all the runs in our database.

| Incident Beam Parameters | | | | Measurements | | | | | |
|---|---|---|---|---|---|---|---|---|---|
| Crystal kind | Energy (keV) | Polarization | | Q factor | | Angle of polarization | | Degree of polarization | |
| | | Angle (°) | Degree (%) | Q | Error | Angle (°) | Error (°) | Fraction (%) | Error (%) |
| CdTe | 200 | 0 | 98 | 0.84 | 0.05 | -1.18 | 0.65 | 91.2 | 1.56 |
| | | | 90 | 0.79 | 0.01 | -0.08 | 0.30 | 79.5 | 0.68 |
| | | | 89 | 0.79 | 0.02 | 0 | 0.30 | 79.1 | 0.85 |
| | | | 88 | 0.77 | 0.02 | 0.09 | 0.33 | 76.6 | 0.74 |
| | | | 85 | 0.80 | 0.02 | -0.34 | 0.30 | 76.1 | 0.81 |
| | | | 80 | 0.78 | 0.04 | 0.27 | 0.50 | 70.1 | 1.07 |
| | | 5 | 98 | 0.80 | 0.02 | 4.84 | 0.22 | 88.0 | 0.5 |
| | | 10 | 98 | 0.80 | 0.02 | 9.7 | 0.21 | 87.9 | 0.5 |
| | | 20 | 98 | 0.79 | 0.02 | 19.62 | 0.21 | 86.7 | 0.5 |
| | | 30 | 98 | 0.79 | 0.02 | 29.55 | 0.20 | 86.1 | 0.05 |
| | 300 | 0 | 98 | 0.74 | 0.02 | 0.09 | 0.29 | 90.4 | 0.75 |
| | | 5 | 98 | 0.76 | 0.01 | 4.81 | 0.20 | 91.2 | 0.53 |
| | | 10 | 98 | 0.76 | 0.01 | 9.79 | 0.21 | 91.3 | 0.54 |
| | | 20 | 98 | 0.76 | 0.02 | 19.91 | 0.21 | 90.6 | 0.55 |
| | | 30 | 98 | 0.76 | 0.01 | 29.74 | 0.20 | 91.1 | 0.51 |
| CZT | 200 | 0 | 98 | 0.79 | 0.01 | -0.6 | 0.09 | 86.2 | 0.2 |
| | | | 90 | 0.77 | 0.02 | -0.48 | 0.07 | 77.6 | 0.16 |
| | | | 89 | 0.78 | 0.02 | -0.23 | 0.23 | 77.0 | 0.52 |
| | | | 88 | 0.75 | 0.01 | -0.59 | 0.22 | 74.4 | 0.5 |
| | | | 85 | 0.75 | 0.02 | -0.54 | 0.09 | 72.0 | 0.19 |
| | | | 80 | 0.75 | 0.03 | -0.27 | 0.40 | 67.6 | 0.85 |
| | | 5 | 98 | 0.78 | 0.02 | 4.62 | 0.20 | 85.6 | 0.49 |
| | | 10 | 98 | 0.78 | 0.01 | 9.36 | 0.05 | 85.0 | 0.13 |
| | | 20 | 98 | 0.77 | 0.01 | 19.37 | 0.18 | 84.3 | 0.43 |
| | | 30 | 98 | 0.77 | 0.01 | 29.66 | 0.18 | 84.3 | 0.44 |
| | 300 | 0 | 98 | 0.73 | 0.03 | -1.07 | 0.37 | 88.1 | 0.95 |
| | | 5 | 98 | 0.73 | 0.01 | 4.65 | 0.15 | 88.2 | 0.53 |
| | | 10 | 98 | 0.72 | 0.01 | 9.93 | 0.22 | 87.6 | 0.55 |
| | | 20 | 98 | 0.72 | 0.01 | 19.73 | 0.21 | 87.3 | 0.53 |
| | | 30 | 98 | 0.73 | 0.01 | 30.05 | 0.19 | 88.5 | 0.5 |

Table 2: Summary of the data configuration taken at ESRF ID15A for Caliste-256, corresponding to our data analysis results.

Figure 12a shows the modulation curves obtained at 200 keV and 300 keV. The modulation factor at both energies is remarkably similar as can be seen by superimposing them on the same plot. $Q$ values of $0.79 \pm 0.02$ and $0.76 \pm 0.01$ are found respectively. In Figure 12b, we illustrate the effect of the angle of polarization on the modulation curve displacement at 200 keV. In this case the angles of polarization given by the fitting procedures are $4.84 \pm 0.22$ and $29.55 \pm 0.20$ for an expected value of 5° and 30° respectively. The corresponding $Q$ factors are $0.80 \pm 0.02$ and $0.79 \pm 0.02$ respectively.

Comparing the 1 mm thick CdTe with the 2 mm CZT based Caliste-256 at 200 keV, with 98% polarization and 0° polarization angle, we find $Q$ to be $0.84 \pm 0.05$ and $0.79 \pm 0.01$ respectively. Looking at the combined data in all conditions, we conclude that within the error bars, both CdTe and CZT behave the same whatever the energy. This is what we would expect due to the fact that we are selecting almost the same subset of Compton

scatters – those around 90°. Of course, the 2 mm thick CZT has the key advantage of a higher efficiency than the 1 mm thick CdTe, which increases the sensitivity but not the figure of merit.

Without any energy selection in our dataset, the modulation factor Q is found to be 0.47 ± 0.003 and 0.41 ± 0.003 for CdTe and CZT modules respectively. These results illustrate the advantage of the fine energy resolution of Caliste-256 to maximize the figure of merit, admitting the efficiency is affected. When compared with other CdTe-based single layer and planar polarimeters, we obtain a higher Q factor (Curado da Silva et al. 2008). This is explained by Caliste-256 prototypes fine pitch, which allows better modulation resolution (Lei et al. 1997, Caroli et al. 2005), and the excellent energy resolution, which allows an optimal angular selection of Compton double events.

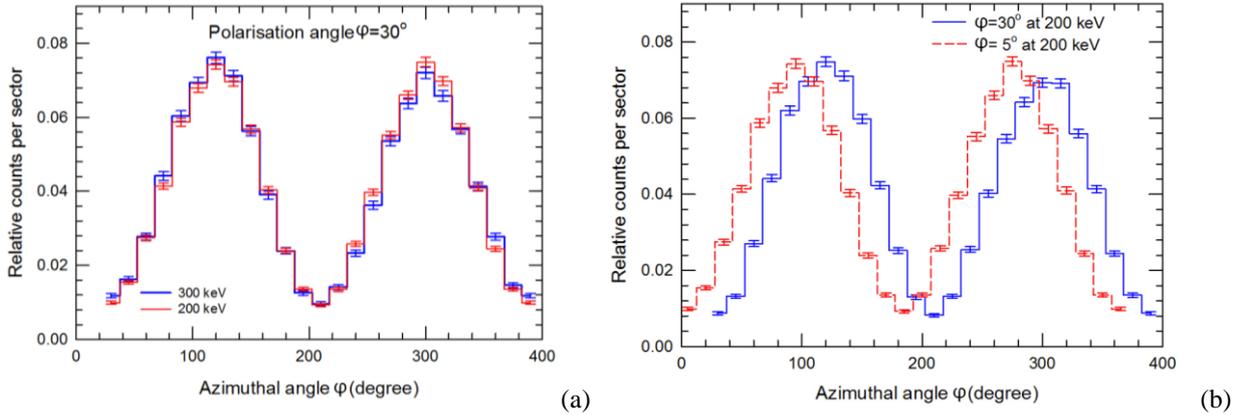

Figure 12. (a) Angular azimuthal distribution of Compton scattered photons (histograms) for a 200 and 300 keV incident beam, angle of polarization $\varphi = 30°$, 98% polarization degree. Scattering angles close to 90° are discriminated by selecting the Compton scattered photon energies from 139 to 148 keV and from 181 to 197 keV for a beam energy of 200 keV and 300 keV respectively; (b) Angular azimuthal distribution of Compton scattered photons (histograms) for two different angles of polarization with a 200 keV incident beam, 98% polarization degree. Scattering angles close to 90° are discriminated by selecting the Compton scattered photon energies from 139 to 148 keV.

In Table 2, one can see that the fitted angles of polarization differ by less than 1° from the nominal angle of polarization of the beam. Figure 13 shows the linear correlation ($\varphi_{obs} = a + b\, \varphi_{beam}$) between the incident beam polarization direction $\varphi_{beam}$ and the measured polarization direction $\varphi_{obs}$. The offset $a$ is less than 1° (-0.5 ± 0.5), which is fully consistent with the alignment uncertainty of the system. The slope $b$ is equal to unity within the uncertainty, (1.00 ± 0.04), which shows that despite the square shape of the pixels, there is no bias in the determination of the beam polarization angle. At all incident energies, the angular results are close to $\varphi_{obs} = \varphi_{beam}$. The small differences can moreover be explained by a small deviation of the beam incidence direction from the perpendicular to the detector surface. The fitting angle of polarization of a simulated modulation curve (with 200 keV, 98% polarized, 0° of angle of polarization) where the detector surface is strictly orthogonal to the beam, is -0.08°, whereas with a $\delta x = 1.5°$ detector surface tilt defined as Figure 11a, the fitted angle of polarization is -0.8°.

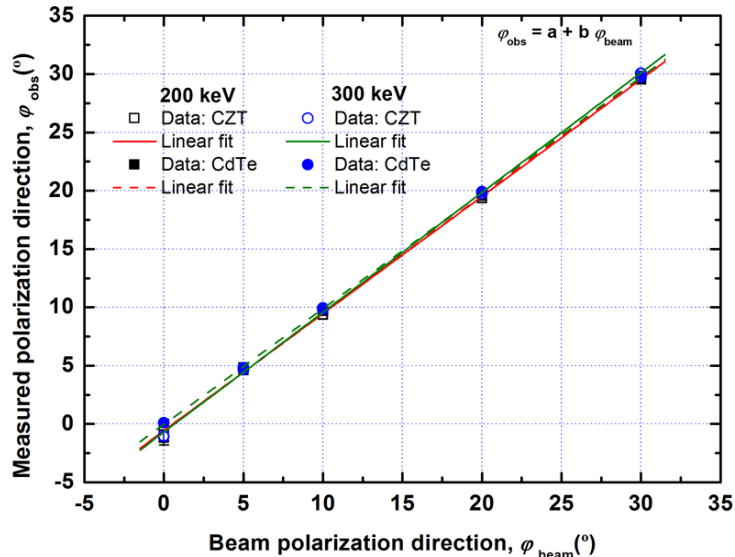

Figure 13. The observed polarization angle ($\varphi_{obs}$) as a function of the ESRF beam polarization angle ($\varphi_{beam}$) at 200 keV and 300 keV, for both CdTe and CZT Caliste modules.

Caliste's performance in resolving the degree of polarization of incoming radiation has been assessed by varying the polarization degree of the ID15A beam from its maximum value (98%) to 80% which is the lowest value achievable, due to beam flux reasons. Previous polarimetric analyses (simulations and measurements) have addressed the polarization degree (Curado da Silva et al. 2012) but in none of these was the instrument resolution to measure this parameter estimated. Figure 14 shows the linear correlation ($P_{obs} = a + b\,P$) between the actual degree of polarization of the beam ($P$) and that measured ($P_{obs}$). The slope $b$, being close to 1 (1.05 for CdTe and 1.15 for CZT), shows that both Caliste modules exhibit a good sensitivity over the full measured range. The $Q$ factor and the associated error bars for polarization levels that differ only by 2% is practically indistinguishable, as the polarization region between 88% and 90% illustrates. Both detectors show an excellent potential to distinguish between polarization degrees different by less than 5%, which is an acceptable performance for studying the different physical processes that generate polarized emissions levels (e.g astrophysical synchrotron about 65% to 80%, magnetic photon splitting: about 20% to 30%, Bremsstrahlung radiation up to 80% (Gluckstern and Hull 1953, Baring et al. 1995, Skibo et al. 1994, Lei et al. 1997).

On the same figure 14 are shown the simulated values. They are fully matching for both Caliste modules, demonstrating that the sensitivity to polarization is not dependent on the thickness and absorption properties within the range tested here. The same conclusion can be drawn within errors from the experimental measurements. The difference between the experimental and simulated curves, which is relatively small, can be probably traced to several reasons. One is that the simulation does not take into account the detector energy resolution and the Doppler broadening, which if done would inevitably blur the polarization information. It might also be that despite this careful energy selection, there is still a small unidentified background component, homogeneous in azimuth, which contributes to lowering the apparent polarization fraction. Finally, as seen before, a modest surface tilt has an effect on the modulation curve, with an impact on the fitted parameters. Investigating these possibilities in more detail is beyond the scope of this paper.

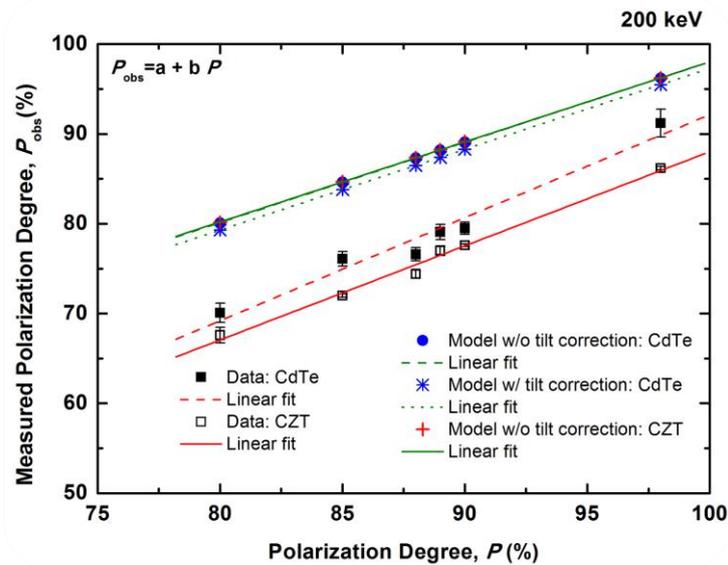

Figure 14. The observed fraction of polarization ($P_{obs}$) as a function of the ESRF beam fraction of polarization ($P$) at 200 keV for both CdTe and CZT Caliste module. Two corresponding simulations at 200 keV for CdTe and CZT where the incident is strictly orthogonal to the beam detector (without correction) and one simulation with the CdTe with the surface detector slightly tilted by 1.5 degree (with correction).

## 8 Conclusions

Through a set of experiments performed at the ESRF, two Caliste-256 prototypes (CdZnTe and CdTe) have been tested in a configuration for determining the system performance when used as a polarization detector at 200 keV and 300 keV. Taking advantage of the excellent energy resolution of these devices, it is possible to perform a simple energy selection to extract the Compton events scattering close to 90° into the detector plane, where the modulation is the highest. Combined with the small size of Caliste-256 pixels, with respect to the interaction lengths at these energies, we have been capable to measure Q factors up to ~ 0.8, much higher than those achieved with the previous types of Cd(Zn)Te detectors. We have also shown that the beam polarization parameters can be measured with an accuracy better than 1° for the polarization angle and better than 5% for the fraction of polarization. These results compare extremely well with a relatively simple analytical simulation model that we have developed, and which moreover, has enabled us to accounts for small geometrical setting effects seen in the data. Of course, a drastic selection in energy to extract the best events causes the efficiency to get lower, which hopefully could be compensated by the use of focusing optics and/or long exposure times, in a science payload. Eventually, the polarization capability of Caliste-256 is used without any compromise in the same acquisition mode than standard time resolved imaging spectroscopy.

The work presented in this paper demonstrates the excellent potential of Caliste-256 modules for being used as a detector for future missions with spectrometric capabilities in the hard X-ray domain.